\documentclass[twocolumn,showpacs,preprintnumbers,a4,prb]{revtex4}
\usepackage{graphicx}
\usepackage{dcolumn}
\usepackage{amsmath}
\usepackage{amssymb}
\usepackage{bm}

\begin{document}

\title{Manipulating superconductivity through the domain structure of a ferromagnet: experimental aspects and theoretical implications}
\author{D. Stamopoulos,\footnote[3]{Author to whom correspondence
should be addressed (densta@ims.demokritos.gr)} and M. Pissas}

\affiliation{Institute of Materials Science, NCSR "Demokritos",
153-10, Aghia Paraskevi, Athens, Greece.}
\date{\today}

\begin{abstract}

In the present work we study experimentally the influence that the
domain structure of a ferromagnet (FM) has on the properties of a
superconductor (SC) in bilayers and multilayers of
La$_{0.60}$Ca$_{0.40}$MnO$_{3}$/Nb and FePt/Nb proximity hybrids.
Specific experimental protocols that were employed in the
performed magnetization measurements enabled us to directly
uncover a generic property of FM/SC hybrids: in the absence of an
external magnetic field, the multidomain structure of the FM
promotes the nucleation of superconductivity, while its monodomain
state strongly suppresses it. Our experimental findings support
recent theoretical studies [A.I. Buzdin and A.S. Mel'nikov, Phys.
Rev. B, 67, 020503(R) (2003), T. Champel and M. Eschrig, Phys.
Rev. B, 71, 220506(R) (2005)] proposing that when an inhomogeneous
exchange field is offered by the FM to the SC the superconducting
pairs are not susceptible to pair-breaking. In contrast, when
magnetic homogeneity is restored in the FM the SC's properties are
strongly suppressed.

\end{abstract}


\maketitle

Today, in solid state physics the basic mechanisms that govern the
fundamental structural, electronic and magnetic physical processes
of pure materials have been widely studied and in most cases well
understood.\cite{Tinkham96,Cullity72} Since the potentiality of
plain materials is restricted by nature, it is the exploration of
artificial hybrids that in recent years has attracted great
interest due to the innovative properties that they could exhibit.
Such a general category of hybrids, which is the subject of the
present work, refers to the combination of ferromagnets (FMs) and
superconductors (SCs).\cite{BuzdinReview,BergeretReview}

In the field of theory it has been proposed that in FM/SC bilayers
near domain walls the destruction of Cooper pairs by the exchange
field is minimized so that in these regimes superconductivity may
be promoted.\cite{Buzdin84} This theoretical concept was based on
the fact that when superconducting pairs (which have a spatial
separation of the coherence length $\xi^{SC}(T)$) are subjected to
an inhomogeneous magnetization of the FM the average exchange
field that they experience over $\xi^{SC}(T)$ could be strongly
reduced compared to the case when the magnetization is
homogeneous. Thus, the magnetic inhomogeneity of the FM
effectively leads to a minimized pair-breaking effect. Since this
condition is fulfilled near domain walls\cite{Buzdin84} it is
expected that in FM/SC hybrids the nucleation of superconductivity
should be promoted when the FM is in a multidomain rather than in
a monodomain state.\cite{Buzdin03R,Buzdin03,Eschrig05,Eschrig05L}
Experimentally, the multidomain state, that is inherent in all FMs
above the Curie critical temperature T$_c^{FM}$, for
T$<$T$_c^{FM}$ is realized near the coercive field where
m$_{FM}=0$. Only few works reported on the experimental
observation of this expectation. J. Aarts, A.I. Buzdin and
coworkers observed sharp drops in the resistance at the coercive
field of appropriately patterned Ni$_{80}$Fe$_{20}$/Nb
bilayers.\cite{Rusanov04} Very recently, V.V. Moshchalkov and
colleagues also observed domain wall assisted superconductivity in
Co-Pd/Nb/Co-Pd trilayer
hybrids.\cite{Moshchalkov05,MoshchalkovNature}

In this work we present magnetization data for
La$_{0.60}$Ca$_{0.40}$MnO$_{3}$/Nb and FePt/Nb bilayers (BLs) and
multilayers (MLs) hybrids. Nb has been chosen as the specific
low-T$_c$ SC since its properties are extensively studied and thus
safe conclusions may be drawn from a direct comparison when the SC
is in pure form and as a constituent of a hybrid. We investigated
different FM materials (La$_{0.60}$Ca$_{0.40}$MnO$_{3}$ and FePt)
and also different structures (BLs and MLs) in order to check the
possible generic character of the obtained results. MLs have been
investigated for one additional reason: these structures offer the
opportunity for intense inhomogeneous magnetic states that may be
efficiently controlled even at nanometer range by regulating the
thickness of the layers that comprise the periodic structure. We
stress that the case where inhomogeneous magnetization is
experienced by a SC in FM/SC structures has been studied
intensively in recent theoretical
works.\cite{BuzdinReview,Bergeret01,Eschrig05,Eschrig05L} In our
work special attention has been paid on the influence of the FM's
domain state on the properties of the SC by employing
minor-loops-based magnetization measurements especially designed
for this purpose. {\it Our results clearly reveal that both in the
BLs and MLs the multidomain state (inhomogeneous magnetization) of
the magnetic constituent promotes superconductivity, while as a
monodomain state (homogeneous magnetization) is established
superconductivity is strongly suppressed.} The effect is more
pronounced for the MLs because they are more magnetically
inhomogeneous due to their artificially produced structural
inhomogeneity. The proposed experimental methodology of minor
loops offers the opportunity to study all FM/SC hybrids
irrespectively of their specific structure.

The La$_{0.60}$Ca$_{0.40}$MnO$_{3}$/Nb and FePt/Nb BLs have
thickness $d_{FM}/d_{SC}=50/100$ and $20/100$ respectively, while
the ML/SC hybrids\cite{StamopoulosPRB05} are
[La$_{0.33}$Ca$_{0.67}$MnO$_{3}$/La$_{0.60}$Ca$_{0.40}$MnO$_{3}]_{15}$/Nb
with [$d_{AF}=4/d_{FM}=4$]$_{15}$/$d_{SC}=100$ (all values are in
nm units). Details on the preparation of the laser-ablated
La$_{0.60}$Ca$_{0.40}$MnO$_{3}$ and dc-sputtered Nb may be found
elsewhere.\cite{Moutis01,Stamopoulos05PRB} The FePt layers were
dc-sputtered on oxidized Si substrates and annealed at $600$ C for
obtaining a hard magnetic phase. The MLs have T$_c^{ML}=230$ K.
The Nb films have T$_c^{SC}=8$ K and $7$ K for the
La$_{0.60}$Ca$_{0.40}$MnO$_{3}$/Nb and FePt/Nb BLs respectively,
while in the La$_{0.60}$Ca$_{0.40}$MnO$_{3}$/Nb MLs they exhibit
T$_c^{SC}=8.2$ K. A commercial SQUID (Quantum Design) was used for
the magnetization measurements.

\begin{figure}[tbp] \centering%
\includegraphics[angle=0,width=7.5cm]{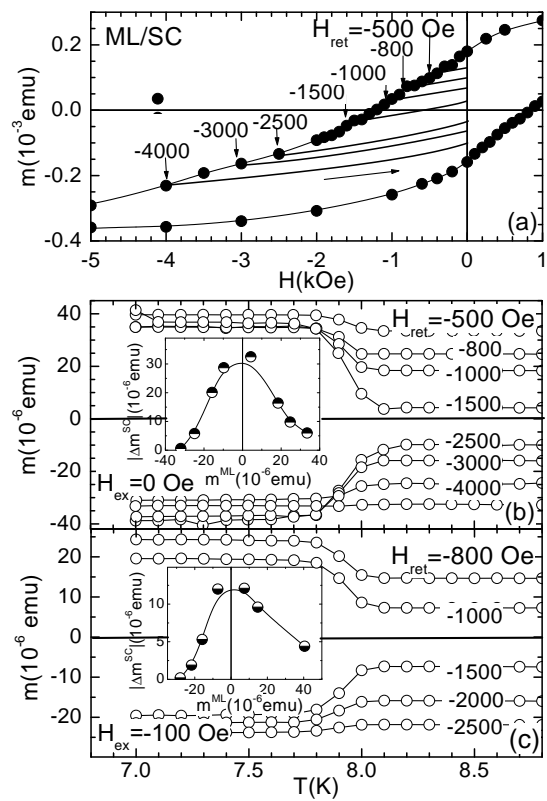}
\caption {(a) Schematic representation of the measuring protocol
performed for a ML/SC hybrid (see text for details). All minor
paths end on a vertical line indicating constant external magnetic
field. The respective magnetization m(T) curves obtained at
external fields $H_{ex}=0$ Oe and $H_{ex}=-100$ Oe are presented
in panels (b) and (c), respectively. Insets present the dependence
of the SC's transition height (absolute value) on the
magnetization of the ML.}
\label{b1}%
\end{figure}

In all measurements the external field was applied parallel to the
hybrids and the magnetization was recorded in the field cooled
(FC) procedure. In order to reveal how the domain state of a FM
influences a SC we propose a specific measuring protocol that is
based on minor magnetization loops. Generally, at constant
temperature $T<T_c^{SC}$ there are two extrinsic parameters that
influence the behaviour of the SC. First, the externally applied
magnetic field H$_{ex}$ and second, the adjacent magnetic
constituent that contributes via both the stray fields that
penetrate the SC and the exchange field that superconducting pairs
experience as soon as they are injected from the SC into the FM.
Thus, our main aim was to find an experimental way to isolate the
influence of the second extrinsic parameter even in case where an
external magnetic field might be present. This may be achieved in
case where the experiments are performed at {\it constant external
magnetic field but for different magnetic states of the FM}. In
order to achieve this goal we performed successive m(T)
measurements when beginning from above the saturation field of the
FM we traced several minor loops by progressively increasing the
field where each new minor loop started. Each one of the minor
loops was accomplished at temperature $T>T_c^{SC}$. After each new
minor loop the external field was kept constant and the
magnetization of the hybrid was recorded as function of
temperature. All the respective m(T) data belonging to a specific
set of measurements {\it were obtained for the same value of the
external magnetic field.} Representative results are shown for a
ML/SC hybrid in Figs.\ref{b1}(a)-(c). In panel (a) we present the
proposed measuring protocol, while in panels (b) and (c) we show
data obtained for H$_{ex}=0$ Oe and H$_{ex}=-100$ Oe respectively.

%------------------------------------------------------------------------
%-------------------------------------------------------------------------

Firstly, we will refer to the case when H$_{ex}=0$ Oe. In this set
of data we initially set $T=10$ K$>T_c^{SC}$ and by starting above
the saturation field of the ML we lower the applied magnetic field
until a characteristic value, that we call the return field
H$_{ret}$, is reached. Then, we start reversing the magnetic field
again until the desired final value is obtained i.e. zero in this
case. At that time we are ready to start the actual m(T)
measurement by lowering the temperature with typical steps of
$50-100$ mK until the transition of the SC is completed. When each
measurement is accomplished we set again $T=10$ K$>T_c^{SC}$ and
we trace the next minor loop, having a higher value of the return
field H$_{ret}$, until the external magnetic field is set back to
zero. Then we are ready to perform the next m(T) measurement. The
whole procedure is repeated as is schematically presented in
Fig.\ref{b1}(a). The obtained m(T) curves that are presented in
panel (b) are indeed revelatory. {\it Since these curves were
obtained at H$_{ex}=0$ Oe it is only the presence of the ML that
affects the SC.} We clearly see that when the ML's remanent
magnetization changes direction it is also the SC's magnetization
that follows (switching effect). These experiments confirm the
results that have been presented very recently in
Ref.\onlinecite{StamopoulosPRB05}. The switching effect survives
even when a non-zero external field is applied as it is shown in
panel (c). The presented data refer to the case when the external
field was set to $-100$ Oe. Since the switching effect should
occur when the ML reverses its magnetic state we expected that
this process should occur at a lower value of the return field
H$_{ret}$ for the data obtained at H$_{ex}=-100$ Oe when compared
to the data obtained at H$_{ex}=0$ Oe. Indeed, this behaviour is
clearly observed in the data presented in Figs.\ref{b1}(b)-(c).
While for the data obtained for H$_{ex}=-100$ Oe the SC switches
its magnetization when the return field is H$_{ret}\simeq -1250$
Oe, in the data where H$_{ex}=0$ Oe the same process occurs at
H$_{ret}\simeq -1700$ Oe. {\it These results indicate that in such
systems we may use the specific state of the magnetic constituent
as an efficient control parameter for the manipulation of the SC's
behaviour.}

%------------------------------------------------------------------------
\begin{figure}[tbp] \centering%
\includegraphics[angle=0,width=7.5cm]{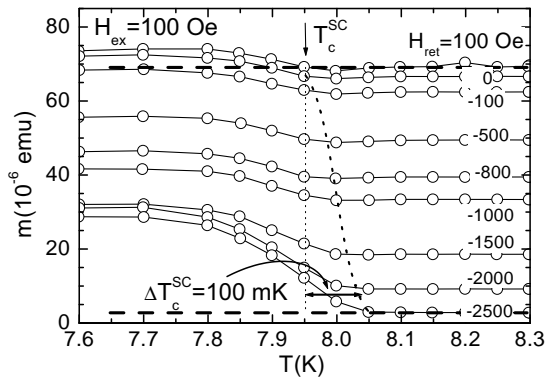}
\caption {Detailed m(T) curves for many magnetization paths of the
ML. Despite the fact that the external field is constant
H$_{ex}=100$ Oe we observe that the SC's transition temperature is
lowered as the magnetization of the ML increases.}
\label{b2}%
\end{figure}%
%-------------------------------------------------------------------------

As already discussed in the introduction in recent years the study
of hybrid structures comprised of FMs and SCs have attracted the
interest of experimentalists
\cite{Rusanov04,Moshchalkov05,StamopoulosPRB05} not only due to
the promising applications that such devices could find in the
near future but also due to the vast theoretical work
\cite{Buzdin84,Bergeret01,Bergeret05} that was done and still
needs experimental feedback. An important experimental outcome of
the present work that is compatible with current theoretical
propositions and could offer significant information for the
further experimental and theoretical examination of FM/SC hybrids
is related to the influence that the domain structure of the FM
has on the SC. In Fig.\ref{b1}(b) we clearly see that the height
of the SC's transition is enhanced significantly as the ML
approaches the state of almost zero bulk magnetization where a
multidomain configuration is acquired. This may be clearly
observed in the insets of panels (b) and (c) where presented is
the variation of the absolute value of the SC's transition height
on the magnetization of the ML. Our results clearly support the
recent theoretical proposals of
Refs.\onlinecite{Buzdin03R,Buzdin03,Eschrig05,Eschrig05L}. Until
now only one experimental report had revealed this effect by
transport measurements in Ni$_{80}$Fe$_{20}$/Nb
bilayers.\cite{Rusanov04} {\it Our magnetization data clearly show
that the multidomain magnetic configuration that the ML acquires
near zero magnetization offers the opportunity to the
superconducting pairs to sample different directions of the
exchange field and thus the pair-breaking effect is only
minimum.\cite{Buzdin84,Buzdin03R,Buzdin03} As a consequence the
SC's transition height is strongly enhanced.}

Except for the transition's height it is also the transition
temperature of the SC which may be affected by the domain
configuration of the FM ingredient. Figure \ref{b2} presents
analytic m(T) curves when many minor paths were traced for
positive value of the constant external field $H_{ex}=100$ Oe. We
clearly see that the maximum T$_c^{SC}$ is observed for
m$_{ML}\approx 0$. When m$_{ML}$ increases the critical
temperature of the SC is shifted to lower values with $\Delta
T_c^{SC}\approx 100$ mK. In addition, we see again that the SC's
transition height is strongly reduced as a monodomain magnetic
state is restored.

%------------------------------------------------------------------------
\begin{figure}[tbp] \centering%
\includegraphics[angle=0,width=8cm]{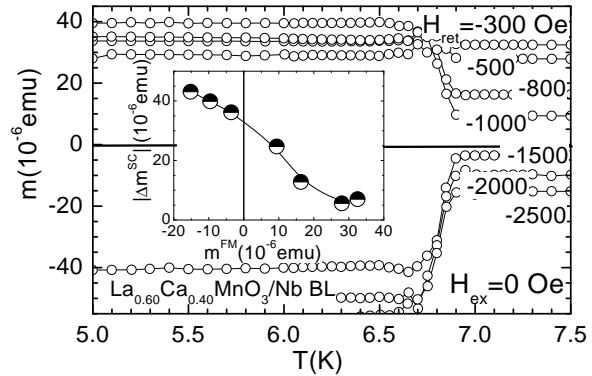}
\caption {Magnetization m(T) curves performed according to the
proposed experimental protocol (see text for details) for a
La$_{0.60}$Ca$_{0.40}$MnO$_{3}$/Nb BL at external magnetic field
$H_{ex}=0$ Oe. Inset presents the dependence of the SC's
transition height (absolute value) on the magnetization of the
La$_{0.60}$Ca$_{0.40}$MnO$_{3}$ layer.}
\label{b3}%
\end{figure}%
%-------------------------------------------------------------------------

\begin{figure}[tbp] \centering%
\includegraphics[angle=0,width=8cm]{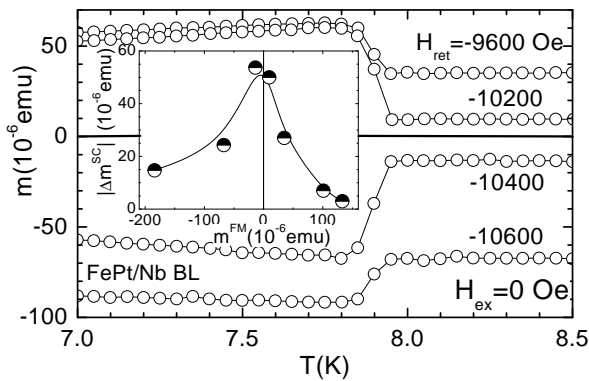}
\caption {Magnetization m(T) curves performed according to the
proposed experimental protocol (see text for details) for a
FePt/Nb BL at external magnetic field $H_{ex}=0$ Oe. Inset
presents the dependence of the SC's transition height (absolute
value) on the magnetization of the FePt layer.}
\label{b4}%
\end{figure}

The same qualitative results were obtained in FM/SC BLs.
Representative measurements for La$_{0.60}$Ca$_{0.40}$MnO$_{3}$/Nb
and FePt/Nb BLs are shown in Fig.\ref{b3} and Fig.\ref{b4},
respectively (the preparative minor loops as the ones presented in
Fig.\ref{b1}(a) for the ML/SC hybrid are not shown). In
Fig.\ref{b3} we clearly see that for external field $H_{ex}=0$ Oe
the switching effect is present and the transition's height
strongly depends on the domain state of the adjacent FM layer as
was observed for the ML/SC hybrids. This may be seen in its inset
where presented is the dependence of the SC's transition height
(absolute value) on the magnetization of the
La$_{0.60}$Ca$_{0.40}$MnO$_{3}$ layer. The respective data
obtained for a FePt/Nb BL are presented in Fig.\ref{b4} for
various values of the return field, $H_{ret}$. Once again, we see
that for $H_{ex}=0$ Oe the transition's height of the SC strongly
depends on the domain structure of the FePt layer. This is clearly
presented in the inset where the SC's transition height (absolute
value) is almost diminished as the FePt layer acquires a
monodomain magnetic structure. Here we should stress a strong
difference between the ML/SC and FM/SC BLs that shows up when an
external magnetic field is applied. In ML/SC hybrids the domain
state of the magnetic constituent still controls the SC's
transition height as may be clearly seen in Fig.\ref{b1}(c). In
contrast, in FM/SC BLs when an external field is applied the SC's
transition height is almost insensitive to the domain state of the
FM layer (results not shown). We attribute this difference to the
more inhomogeneous magnetic state that a ML exhibits (due to its
specific structure) when compared to a single FM layer.\cite{note}

Apart from the mechanism of the exchange interaction we would like
to discuss briefly the influence that the electromagnetic
mechanism might have on superconducting pairs since the FM
generates stray fields that enter the SC alongside their
interface.\cite{BuzdinReview,Buzdin84,Buzdin03R,Buzdin03} The SC's
transition height is maximized near the FM's zero magnetization
where the stray fields that accompany the respective magnetic
domains should be randomly distributed. As a result the average
{\it macroscopic} stray field experienced by the adjacent SC film
should be almost zero and thus couldn't motivate {\it bulk}
superconductivity. However, the local contribution of the stray
fields in the {\it mesoscopic} range can't be neglected entirely.
In this sense, since our results indicate that superconductivity
is enhanced when the domain walls are maximized, we may assume
that the observed effect could be partly motivated by the
electromagnetic mechanism as observed very recently in the
trilayered Co-Pd/Nb/Co-Pd structures studied in
Ref.\onlinecite{Moshchalkov05}.

%------------------------------------------------------------------------
%-------------------------------------------------------------------------

Summarizing, in this work we presented magnetization data for BL
and ML hybrids consisting of La$_{0.60}$Ca$_{0.40}$MnO$_{3}$ and
FePt combined with low-T$_c$ SC Nb. By employing specific
measuring protocols we isolated the interplay between the domain
configuration of a FM and the nucleation of superconductivity in
an adjacent SC: the inhomogeneous exchange field related to a
multidomain magnetic state clearly promotes the nucleation of
superconductivity, while as homogeneity is restored and a
monodomain magnetic state is established superconductivity is
strongly suppressed. The effect is more pronounced for the MLs
when compared to the BLs due to the more inhomogeneous
magnetization that they exhibit intrinsically owing to their
specific structure. We speculate that our experimental
observations should be inherent in all FM/SC hybrids. We hope that
our study will trigger farther experimental and theoretical works
on the possible existence of such phenomena in relative hybrid
structures.

\begin{acknowledgments}
Dr. N. Moutis and Dr. E. Manios are acknowledged for valuable help
during the preparation of samples. Mr. P. Tabourlos should be
warmly acknowledged for valuable technical assistance.

\end{acknowledgments}

\pagebreak


\begin{references}
\bibitem{note}Here we have to recall that the MLs are comprised of adjacent AF
and FM layers of thickness only $d_{AF}=d_{FM}=4$ nm. In contrast,
the single FM layer that is used in the BLs has thickness $50$ nm
and $20$ nm for the La$_{0.60}$Ca$_{0.40}$MnO$_{3}$/Nb and FePt/Nb
structures respectively. Since the maximum size of each magnetic
domain, at least in the direction along the thickness of the
sample, is limited by the thickness of each respective layer, it
is obvious that the MLs are more magnetically inhomogeneous when
compared to the BLs.
\end{references}
\end{document}